# The impact of asset pre-selection for a cardinality constrained index tracking portfolio with optional enhancement

**N. Meade[1], C.A. Valle[2] and J.E. Beasley[3]**

[1]**Imperial College Business School, Imperial College, London SW7 2AZ, United Kingdom**

**n.meade@imperial.ac.uk**

[2]**Depto. de Ciência da Computação, Universidade Federal de Minas Gerais, Belo Horizonte - MG, Brazil.**

**arbex@dcc.ufmg.br**

[3]**Mathematics, Brunel University, Uxbridge UB8 3PH, United Kingdom**

**john.beasley@brunel.ac.uk**

# The impact of asset pre-selection for a cardinality constrained index tracking portfolio with optional enhancement

**Abstract**

Index trackers are important passive investments offering the return and risk of the market encapsulated by the index, the largest US index tracker was valued at $900 billion in early 2026. The problem of selecting a cardinality constrained index tracker has attracted much interest in the literature. Using a two-stage approach of asset selection followed by estimation on S&P 500 data, we explore the role of cardinality constraints in determining the effectiveness of the tracker's reproduction of market return and risk. We compare eight pre-selection procedures: forward selection or backward elimination; implemented using ordinary least squares or least absolute deviation regression; with or without a regression constant. We show experimentally that out-of-sample tracking errors decrease according to $1/\sqrt{cardinality}$ and out-of-sample tracking error, transaction volume and return-risk ratios all improve as the cardinality constraint is relaxed. By contrast for enhanced returns, cardinalities of the order 10 to 20 are most effective.

**Data Availability:** The S&P 500 data used from 3/1/2005 to 29/12/2023 is available on https://github.com/cristianoarbex/cardinalityPortfolioData.

**Conflict of Interest:** There are no conflicts of interest for any of the authors.

**Funding**: No funding was received.

## 1. Introduction

Investment in the stock market can be divided into two categories, active and passive. An active fund manager seeks to add value by their choice of assets and by the timing of asset purchases and sales. A passive fund manager endeavours to reproduce the return on the market as a whole. A common passive investment is a fund that tracks a market index that summarises a market's performance. An example market index is the S&P 500 which encapsulates the value of the largest US corporations. Passive investment has become increasingly popular over the last decade, according to research cited by the Financial Times (FT Money, 27 July 2024, p3) £77 billion in fees have been lost to active managers in Europe since 2011. Sharpe (1991) discusses the difference between active and passive investment, indicating that "after costs, the return on the average actively managed dollar will be less than the return on the average passively managed dollar." However, Sharpe defines a passive investor as one whose portfolio exactly reflects the market considered by the tracked index. In practice, exact replication of the market index involves monitoring and adjusting a portfolio of many assets (500 for the US S&P, nearly 2000 in the Japanese Topix, 563 in the UK FT All Share). While many passive fund managers do carry out full replication, forming a tracking portfolio with fewer assets than the index is a feasible alternative. Quantitative analysis underlying the selection of index tracking portfolios has been much discussed in the literature (early studies include Rudd (1980), Andrews et al. (1986), Congdon (1987), Nash (1988), Meade and Salkin (1989)). More recently, Silva and de Almeida-Filho (2024) give a comprehensive review of 117 studies dealing with index tracking.

An alternative view of stock market investments is that investment strategies are on a spectrum with subjective judgement determining asset choice and transaction timing at one end and full replication of the index at the other end. Between these extremes there are algorithmic approaches where different objective functions determine asset choice and transaction timing; these may include cardinality constrained index tracking or delivering an enhanced return on an index. Here, we consider the problem of selecting a portfolio to closely reproduce the risk and return of the market index (without or with an enhancement) with a subset of assets chosen from those represented in the index.

Selection of the assets and the estimation of their associated weights in a cardinality constrained index tracking portfolio (CCITP) is a computationally difficult optimisation problem. Until the mid-2010s this problem needed to be addressed using a heuristic procedure. Advances in hardware and solver development since then mean that optimal solutions can now be produced for relatively large cardinalities (increasing over time by an order of magnitude from ten to hundreds) for many asset universes. In much of the literature, the asset selection process is treated as an integral part of the optimisation problem, here we propose to pre-select the assets for a given cardinality using a range of statistical procedures. Once a given set of assets has been identified, the weights on each asset within the portfolio can be found straightforwardly. Due to the computational difficulty of using heuristics to solve the problem associated with jointly choosing the asset set and their weights in the portfolio, many examples in the research literature (especially before 2020) consider portfolios of no more than 10 assets. Using one of the procedures proposed in this paper to pre-select the assets in a cardinality constrained index tracking portfolio removes this problem.

For an index with $M$ constituent assets, our proposed approach is in three stages:

- a choice procedure is used to select the assets for a given cardinality, $n$ $(n < M)$;
- the weights on assets for a cardinality constrained portfolio with $n$ assets are optimised to minimise in-sample tracking error;
- an optimal cardinality is chosen according to an acceptability criterion.

An attraction of this approach is that our formulation allows for an enhancement to the return on the index.

The structure of this paper is as follows. In the next section we review the literature. In the third section we define the problem and introduce the proposed choice procedures. The data set we use is described in the fourth section. In the fifth section, we describe our analysis and discuss the choice of cardinality for index tracking portfolios and for portfolios offering enhanced returns. Our conclusions are given in the sixth section.

## 2. Literature Review

An index-tracking fund is a form of passive investment in a particular market summarised by a published index. Index-tracking funds are available from many providers, a popular form is an index tracking exchange traded fund (ETF), in 2022 there were 77 ETFs tracking the S&P 500 and 11 ETFs tracking the FTSE100. The largest S&P 500 tracking ETF has over US$900 billion in assets under management (as of May 2026).

Meade and Salkin (1989) identify three approaches to constructing an index fund: full replication; stratification and sampling. Typically, full replication has the greatest management costs and the most accurate tracking of the index; sampling means choosing a subset of assets to reproduce index behaviour at a lower management cost but less accurate tracking. Stratification means constraining a sampled portfolio to reproduce the different sectors within the index. The sampling approach has received by far the greatest coverage in the literature.

Approaches to the problem are heuristic-based or mathematical programming-based or a hybrid of the two approaches. In their very recent review of solution procedures for the index tracking problem, Silva and de Almeida-Filho (2024) listed 21 heuristic approaches in 53 studies; 8 types of mathematical programming approaches in 84 studies; the metrics used in the studies they reviewed varied widely, including computation time (11%) and (root) mean square error (11%); 41% of studies included a cardinality constraint. The indices tracked in these studies are not identified, only the data sources. Dhingra et al. (2026) also offer a review of index tracking literature, where they contrast different methodologies: mathematical programming and decision-analytic; statistical and econometric;

machine-learning. Given the comprehensive reviews of Silva and de Almeida-Filho (2024) and Dhingra et al. (2026), we refer the interested reader to their work. In the first subsection, we discuss: the determinants of the difficulty of the CCITP and give examples of the range of problem sizes covered in the literature. In the second subsection, we review approaches to enhanced index tracking where returns in excess of the index are sought.

*2.1 Examples showing the sizes of problems studied*

The difficulty of the CCITP problem is largely determined by the number of assets in the index tracked, *M*, and the cardinality, *n*. Imposing a cardinality constraint on portfolio selection means choosing *n* assets from the universe of *M* assets giving $\binom{M}{n}$ possible choices. Selecting a cardinality constrained tracking portfolio for the DAX 40 is 'easier' than selecting one for the Russell 3000 growth index since $\binom{40}{n} \ll \binom{3000}{n}$. Similarly, selecting a portfolio of 10 assets to track the S&P 500 is very much easier than selecting a portfolio of 250 assets since $\binom{500}{10} \ll \binom{500}{250}$. For this reason, the cardinality used in studies is often quite small, for example: Beasley et al (2003) demonstrated an evolutionary heuristic algorithm for the CCITP using a cardinality of 10 ($31 \leq M \leq 225$); Xu et al. (2015) used a non-monotone projected gradient method as a local minimiser with cardinality of $n = 5,\ldots,10$ ($31 \leq M \leq 3000$). Garcıa et al. (2018) compared a genetic algorithm (GA) approach with tabu search using a maximum cardinality of 10 ($31 \leq M \leq 225$). Rubio-García et al. (2024), used a hybrid simulated annealing algorithm to find quasi-optimal results for cardinalities between 10 and 30 assets ($M = 433$). A contrasting approach is implemented by Jansen and Dijk (2002) who used quadratic programming iteratively ($M = 500$), with no cardinality constraint in the first iteration, and in subsequent iterations removed sets of the lowest weighted assets, until the required cardinality was reached. Coleman et al. (2006) proposed a graduated non-convexity method, which they compared with a GA and the Jansen and Dijk (2002) approach ($M = 500$). Silva et al. (2024) used the greedy randomised adaptive search procedure (GRASP) metaheuristic, comparing their results with a GA approach ($31 \leq M \leq 225$). Xu et al. (2026) make use of an asset dependency structure based upon return correlations which results in a mixed-integer quadratic programming problem. They give computational results relating to two Chinese stock market indices ($30 \leq M \leq 500$). An example of a computer intensive solution method considering cardinalities up to 100 assets is Palmer et al. (2022), who use quantum computing ($100 \leq M \leq 500$). Considering cardinality constraints of 10%, 20% and 30% of the available universe of S&P 500 stocks, Zhang and De Smedt (2026) compare 'end-to-end' approaches to solving the CCITP problem using Long Short-Term Memory (LSTM) networks in conjunction with different Softmax functions ($M = 481$).

In several studies the proposed algorithms consider the selection of *n* assets separately from the computation of the weights. Dose and Cincotti (2005) use time-series cluster analysis to select assets and then use stochastic optimisation to determine asset weights in CCITPs (M = 500). For a range of cardinalities up to n = M = 15 assets (from the PHLX Oil Service Sector) Fernández-Lorenzo et al. (2021) used 'a *k*-step pruning' procedure in conjunction with convex optimisation to guide asset selection. With a maximum cardinality of 10, Torrubiano and Alberto (2009) used a GA to select the assets and a quadratic program to compute asset weights ($31 \leq M \leq 225$). As a contrast, Mutunge and Haugland (2018) considered values of *n* up to 100 ($500 \leq M \leq 3000$). As *n* is increased, they exploit the structure of the covariance matrix of asset returns, using a greedy heuristic to extend the current portfolio by a single asset at each iteration so as to minimise tracking error.

*2.2 Studies in enhanced index tracking*

An early proposal relating to enhanced index tracking was made by Beasley et al (2003), who suggested investors would be attracted by a portfolio offering a return in excess of the index return while minimising tracking error. Guastaroba et al. (2020) define the enhanced index tracking problem as the determination of an optimal portfolio of assets with the bi-objective of maximising the excess return of the portfolio above a benchmark and minimising the tracking error. As an alternative to a root mean square tracking error, they propose a class of bi-criteria optimisation models where risk is measured using a weighted combination of multiple conditional value at risk measures, allowing a more detailed investigation of risk aversion. Sadeghi et al. (2026) also used a bi-level optimisation approach to the enhanced index tracking problem, finding that mean squared deviation generated better results as compared with other objective functions. Various techniques for the solution of the enhanced index tracking problem have been investigated. In conjunction with the capital asset pricing model, Canakgoz and Beasley (2009) propose a mixed-integer programming approach to the CCITP, using the beta of the selected index tracking portfolio to control enhancement. This approach is also used by Filippi et al. (2016) who propose a heuristic procedure to approximate a set of Pareto optimal solutions. Mezali and Beasley (2013) present a similar approach using quantile regression. Bonomelli et al. (2026) also used quantile regression subject to stochastic dominance constraints. In the spirit of Canakgoz and Beasley (2009), Chavez-Bedoya and Birge (2014) use non-linear programming to solve a nonconvex objective function which includes the correlation between the CCITP and the index plus other risk measures. Using signal processing techniques, Li and Bao (2014) extract multiple time-scale features of asset return series using empirical mode decomposition, an immunity based multi-objective optimisation algorithm is used to develop enhanced index tracking portfolios. Using a probability-based criterion to identify assets with positive momentum, Meade and Beasley (2011) construct portfolios designed to outperform market indices. They find evidence of significant momentum profits (including reasonable transaction costs) in seven out of eleven indices investigated. Looking at financial practice in Australia, Frino et al. (2005) compared the portfolio configuration strategies of five index and three enhanced index equity funds. They found that the enhanced index funds began portfolio rebalancing earlier, and used more patient trading strategies than the index funds, generating higher returns and lower trading costs for the enhanced index funds.

## 3. Problem Definition

A distinctive feature of any portfolio selection problem, including the CCITP problem, is that the selection is based on making use of available/known in-sample data, but implemented on unavailable/unknown out-of-sample data. As a consequence, striving to reduce in-sample tracking error by extra fractions of a percent is unlikely to be rewarded out-of-sample

The approach we propose in this paper is in two stages, firstly we select the assets to be included in the CCITP, secondly, we estimate the asset weights. Separation of these stages permits the use of two well established procedures, regression and quasi-Newton estimation. As far as we are aware this approach has not been presented previously and so our work constitutes a new and distinct contribution to the literature. Our objective is to measure the effect of the cardinality constraint on out-of-sample tracking error for index trackers and enhanced index trackers.

### *3.1 Selection procedures*

The probability density function of the returns on the assets in the index should inform the choice of selection method. Adcock and Meade (2017) proposed a parametric classification procedure which showed that most asset return distributions from several stock markets were best described by a generalised *t* distribution, a small minority were well described by the Normal distribution and only a few were captured by the Laplace distribution. Implementation of the generalised *t* distribution in a regression would involve a bespoke maximum likelihood approach, beyond the scope of our analysis, thus we restrict our analysis to the two readily available regression methods using ordinary least squares (OLS) and least absolute deviation (LAD) objectives. OLS regression assumes a multivariate normal distribution for asset returns, where asset return behaviour is completely described by a mean vector and a covariance matrix, an assumption underlying Markowitz portfolio selection. LAD regression assumes Laplace distributed returns, where the distribution of returns has fatter tails than the normal distribution, this assumption encompasses the 'fat tails' where extreme values of returns (more than three standard deviations) occur more often than is consistent with the normal distribution. Less well known than OLS, LAD regression uses a linear programming approach proposed by Barrodale and Roberts (1973).

Given a universe of $M$ assets, the objective of the selection stage is to identify sets of $n$ assets for $n = 1,..,n_{max}$, where $n_{max} \leq M$, these identifications are summarised by the ($n_{max} \times M$) matrix $\Delta$. The elements of $\Delta$, $\delta_{ni}$, define the membership of a tracking portfolio of $n$ assets.

$$\delta_{ni} = \begin{cases} 1 & if\ asset\ i\ is\ chosen \\ 0 & if\ asset\ i\ is\ not\ chosen \end{cases}$$

where $i = 1,\ldots, M$.

*3.1.1 Forward selection*

In this procedure we rank the assets in ascending order of their power to explain the variability of the index. We use regression sequentially to identify the next most suitable asset to include in the tracking portfolio as its cardinality increases. Once an asset is chosen for a set of cardinality $n$, it is present for all sets of cardinalities greater than $n$, so if $\delta_{nj} = 1$ then $\delta_{sj} = 1$ for $s > n$. The first asset is chosen by offering the best fit in a simple regression, where for OLS, the asset maximises the adjusted $R^2$; for LAD, the asset minimises the mean absolute deviation (MAD). Throughout our analysis, we use regression with a constant denoted by $(\kappa = 1)$ and without, denoted by $(\kappa = 0)$. We include the opportunity to track an enhanced index by adding a target enhancement, λ, to the index return, for straightforward index tracking λ is set to zero. More formally, for cardinality $n = 1$, asset $i$ is chosen, $\delta_{1i} = 1$, to maximise the explanatory power of the regression:

$$ln\left({I_t}/{I_{t-1}}\right) + \lambda = \kappa\alpha_1 + \beta_{1i} ln\left({P_{it}}/{P_{i,t-1}}\right) + \varepsilon_{1t} \quad (1)$$

where $I_t$ is the value of the index at time $t$ and $P_{it}$ is the price of asset $i$ at time $t$. Subsequent assets are chosen for their power in explaining the residual variation in the previous regression, that is variation in index returns unexplained by the assets already chosen. Formally, for $n>1$, the n[th] asset chosen, $\delta_{ni} = 1$, is asset $i$ (where $\delta_{si} = 0 \; \forall s \leq n-1$) which maximises the explanatory power (measured by adjusted $R^2$ for OLS or MAD for LAD) of

$$\varepsilon_{n-1,t} = \kappa\varphi_n + \omega_n ln\left({P_{it}}/{P_{i,t-1}}\right) + \gamma_{nit} \quad (2)$$

As $n$ is increased the residual variation unexplained by the chosen set of assets is updated by the following multiple regression:

$$ln\left({I_t}/{I_{t-1}}\right) + \lambda = \kappa\alpha_n + \sum_{j \; s.t. \; \delta_{nj}=1} \beta_{nj} ln\left({P_{jt}}/{P_{j,t-1}}\right) + \varepsilon_{nt} \quad (3)$$

where $\alpha_n$ and $\beta_{nj}$ are OLS/LAD regression coefficients when the index return is explained by the returns on the $n$ assets already chosen. In OLS, the standard error of the estimate $\beta_{ni}$ is denoted $se(\beta_{ni})$, its relative significance is judged by its t-value $\left|{\beta_{ni}}/{se(\beta_{ni})}\right|$. Assets are chosen by repeating steps (2) and (3) until $n = n_{max}$.

*3.1.2 Backward elimination*

The starting point of this procedure is that all available assets are chosen initially and assets are sequentially removed, one at a time. The full set of asset returns are regressed against the return on the index. The asset with the least explanatory power is eliminated and the returns of the remaining assets are regressed against the return on the index. This sequence is repeated until there is one asset remaining. For OLS, least explanatory power equates to the asset with the minimum $t$-value; for LAD, the variables are standardised before the regression and then the asset with the least explanatory power has the smallest absolute value of its regression coefficient, $|\beta_{ni}|$. The first stage of the procedure reduces the set of eligible assets from the universe of $M$ assets to the maximum cardinality considered, $n_{max}$. The cardinality is then reduced by one at each step, the regression (3) is estimated for cardinality $n$ ($n = n_{max},\ldots,2$ ) and the asset $i$ that contributes least explanatory power is eliminated from the set of eligible assets to form a set of cardinality $n - 1$, once we set $\delta_{ni} = 0$ then $\delta_{si} = 0 \ \forall s < n$.

### *3.1.3 A summary and identification of alternative procedures*

Forward selection and backward elimination both have strengths and weaknesses. Forward selection can be used for any size of universe; however, in the initial stages backward elimination needs more observations than there are assets in the index for a non-degenerate solution. Thus, this procedure may not be viable with newly introduced indices or the Russell indices of 1000 or more assets.

Backward elimination may capture useful correlations between asset returns (and with index returns) which forward selection may fail to capture. For example, suppose backward elimination chooses asset $a$ and asset $b$ for its portfolio of cardinality two, (as together they produce the highest adjusted $R^2$), but since forward selection chose asset $c$ for its single asset portfolio (as it has the highest correlation with the index), forward selection cannot choose both asset $a$ and asset $b$ in its portfolio of cardinality two.

In summary, we investigate the following procedures for choosing assets for values of $n = 1,\ldots,n_{max}$

| | |
|---|---|
| Forward selection using ordinary least squares regression: | FS-OLS(c or n) |
| Backward elimination using ordinary least squares regression: | BE-OLS(c or n) |
| Forward selection using least absolute deviation regression: | FS-LAD(c or n) |
| Backward elimination using least absolute deviation regression: | BE-LAD(c or n) |

where (c or n) means with a **c**onstant ($\kappa = 1$) or **n**o constant ($\kappa = 0$).

### ***3.2 The tracking error objective function***

We use a moving time window, where we iteratively select the tracking portfolio and evaluate its performance. We will use $T_1$ observations to select the tracking portfolio and $T_2$ observations for evaluation. The process is shown in Figure 1.

For convenience in discussing results, $T_1$ and $T_2$ will be identified as either a period, i.e. a number of years or months, or in formulae as the number of observations corresponding to the period. For example, if the observations are daily and $T_1$ is 1 year then in a formula $T_1 \approx 252$, depending on the calendar. Thus, the $K^{th}$ period in the sequence of selection and evaluation will use observations: $[1 + (K - 1)T_2 : T_1 + (K - 1)T_2]$ for portfolio selection and use observations:
$[1 + T_1 + (K - 1)T_2 : T_1 + KT_2]$ for out-of-sample evaluation.

**Figure 1. Representation of the iterative tracking portfolio selection and evaluation cycle**

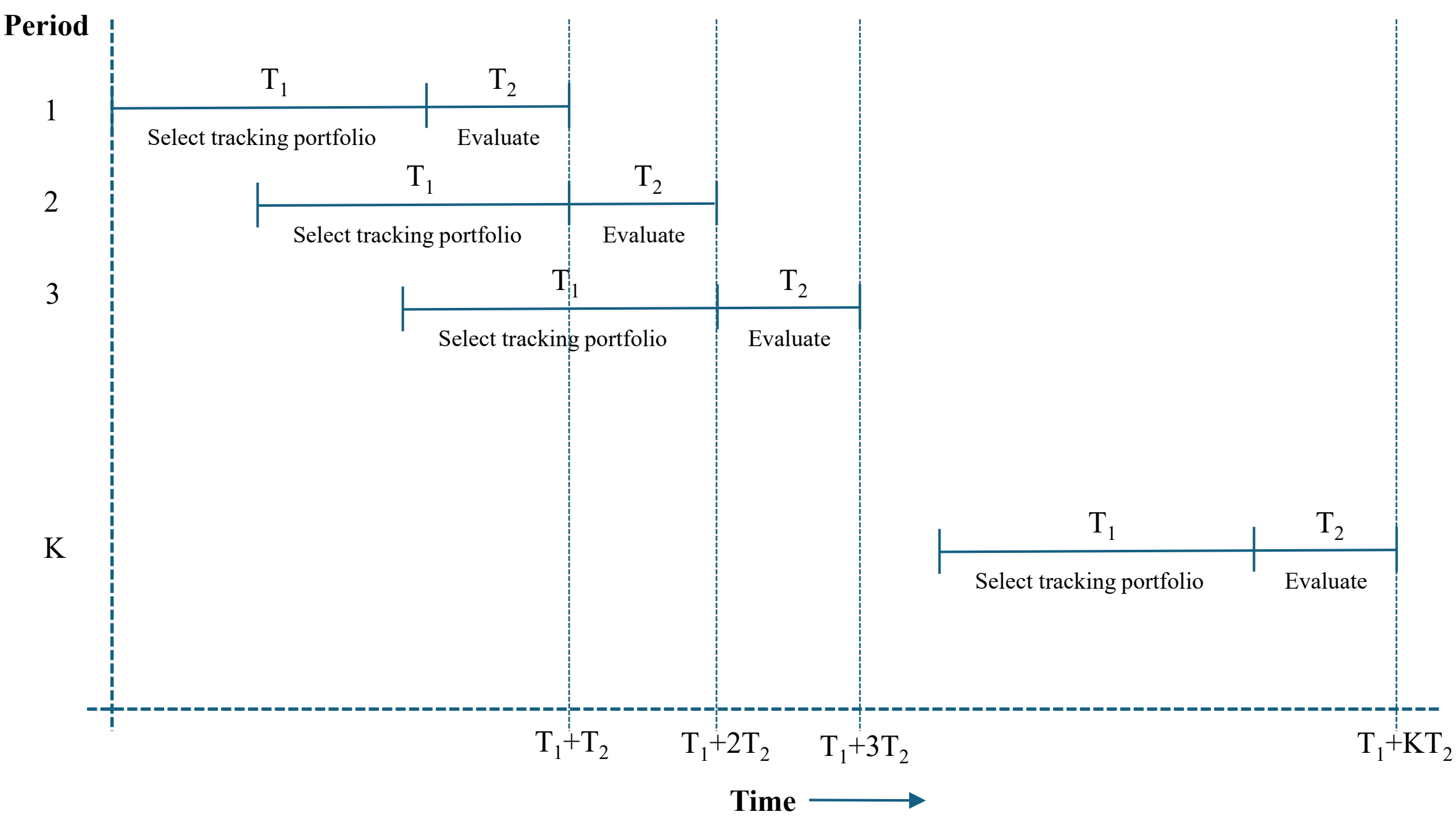

In the portfolio selection stage (in-sample), the asset weights in the tracking portfolio are chosen to minimise an appropriate objective function, as seen below. The tracking portfolio is comprised of $n$ assets, asset $i$ is included in this portfolio if $\delta_{ni} = 1$ and has a weight of $x_i$ where $x_i > 0$ and $\sum_{i:\delta_{ni}=1} x_i = 1$. The weights, $x_i$, are optimised by minimising the tracking error over $T_1$ observations. We focus our analysis on a root mean square tracking error, since this form is most common in the literature and has the same units as asset volatility. Tracking error is defined below:

$$TE = \left( \frac{\sum_{t=t_0+1}^{t_0+m} \left( \left( ln\left( {}^{I_t}/_{I_{t-1}} \right) + \lambda \right) - ln\left( {}^{\sum_i x_i P_{it}}/_{\sum_i x_i P_{i,t-1}} \right) \right)^2}{m} \right)^{1/2}, \quad (4)$$

$m$ is the number of time periods considered, $m = T_1$ for selection and $m = T_2$ for out-of-sample evaluation, $t_0$ is determined by the value of $K$. To ensure that the tracking portfolio reproduces the same mean return as the index, we introduce a penalty function which is the squared difference of the mean returns on the index and the tracking portfolio:

$$Penalty = \left( \frac{\sum_{t=t_0+1}^{t_0+m} \left( \left( ln\left( {}^{I_t}/_{I_{t-1}} \right) + \lambda \right) - ln\left( {}^{\sum_i x_i P_{it}}/_{\sum_i x_i P_{i,t-1}} \right) \right)}{m} \right)^2 \quad (5)$$

To clarify the difference between *TE* (representing the difference in risk between tracking portfolio and index) and *Penalty* (representing the difference in mean return), note that *TE* has the sum of squared terms in its numerator, *Penalty* has the sum of non-squared terms. Thus, if λ=0 and over the period $t_0+1$ to $t_0+m$ the total return from the index is equal to the total return from the portfolio *Penalty* will be zero. However, even if *Penalty* is zero *TE* may be non-zero, *TE* can only be zero if the return from the index is equal to the return from the portfolio in each and every time period *t*. The objective function to be minimised is thus

$$TE^2 + weight \times Penalty \tag{6}$$

In our analysis a search over values of the weight showed that *weight* = 5 achieves equality between the index return and the return on the tracking portfolio. Since this objective is a non-linear function of the $x_i$ variables, for reasons of computational convenience, we use a quasi-Newton method (specifically routine BCONF from the IMSL FORTRAN Maths library, https://help.imsl.com/fortran/fnlmath/current/BCONF.htm) to solve the minimisation.

Moving from period $K$ to $K$+1, the index tracking portfolio is rebalanced (asset weights change from $x_{iK}$ to $x_{i,K+1}$) and transaction costs are incurred. Exact reproduction of transaction costs is complicated depending on the type of financial institution involved, see Collins and Fabozzi (1991). An analysis by Schwarz et al. (2022) demonstrated that the number of shares transacted was the major determinant of cost. For simplicity, we follow Adcock and Meade (1994) and assume that a component of the transaction costs incurred when rebalancing from period $K$ to period $K$+1 is proportional to $\sum_i \lfloor x_{i,K+1} - x_{iK} \rfloor$. For convenience, we will refer to this quantity as the transaction volume.

## 4. Data

Our full data set is daily asset prices for 822 assets which have been members of the S&P 500 Composite Price Index (at some time or another) from 3/1/2005 to 29/12/2023, so nearly two decades of data. For our analyses, we restrict attention at each rebalance only to those assets that are index members at the time of the rebalance, our data set has been extensively and manually curated to ensure this. To facilitate further research, our dataset is publicly available to researchers on https://github.com/cristianoarbex/cardinalityPortfolioData.

The index and its annualised volatility are shown in Figure 2. The index trends upwards throughout the period, with occasional reverses, notably during the 2009 financial crisis and the effects of Covid during 2020 – 2021. The volatility increases during these reverses reflect market uncertainty; these periods are likely to be more challenging for tracking the index.

**Figure 2. Daily values of the S&P 500 Composite Price Index from 3/1/2005 to 29/12/2023 with its annualised volatility.**

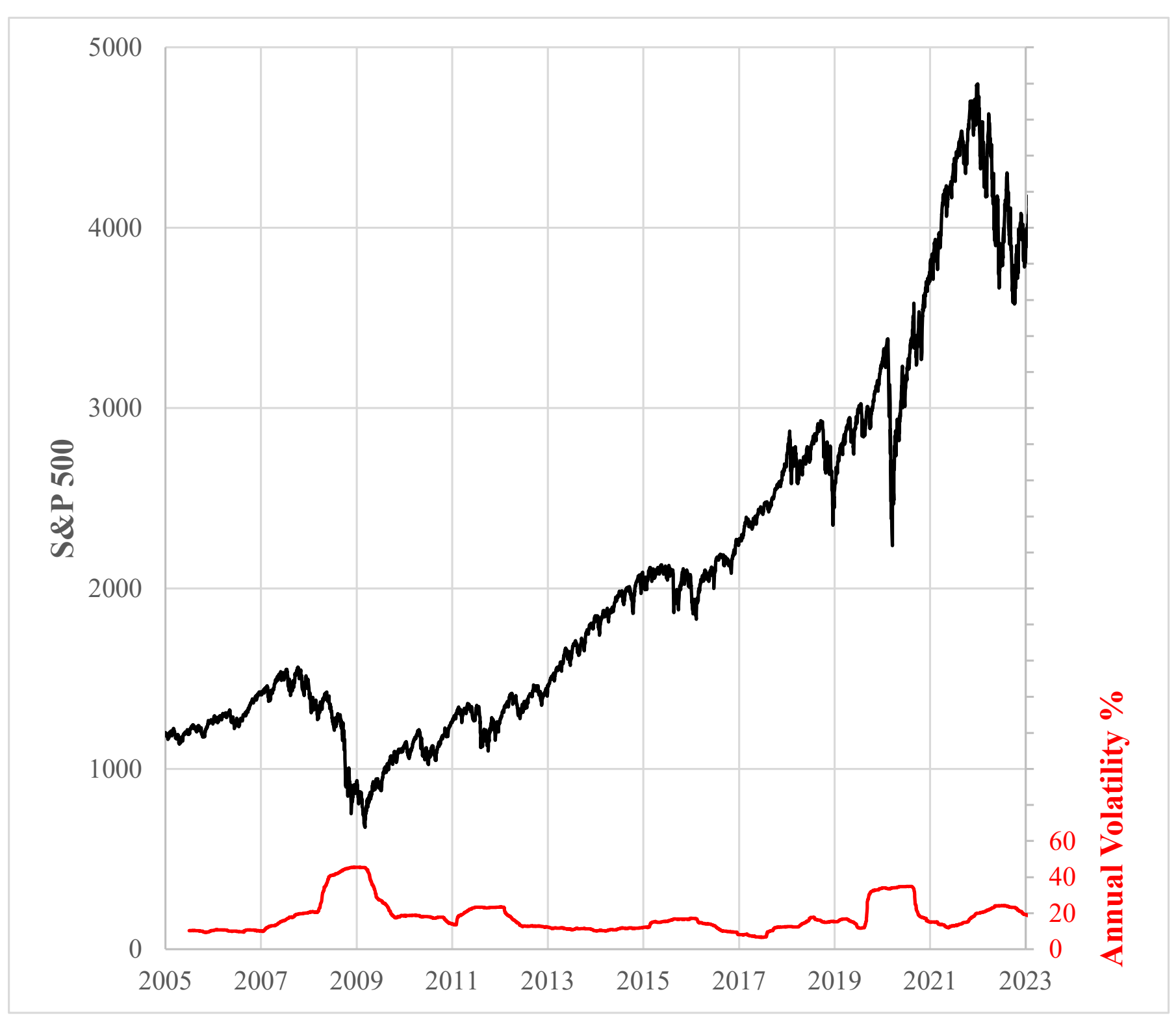

**5. Analysis**

Our analysis is divided into five sections. In section 5.1, we use the full set of daily data for our analysis comparing the performance of the pre-selection procedures over a sequence of tracking portfolio selections and evaluations. Having identified the most effective pre-selection procedure, we focus on using it in the remaining sub-sections. In section 5.2, we devise a model to explain the observed decrease in tracking error as cardinality is increased. To help understand the variation in tracking error from period to period, we relate out-of-sample tracking error to index volatility in section 5.3. We investigate tracking an enhanced index in section 5.4 and carry out a sensitivity analysis to find effective combinations of estimation period, $T_1$, and evaluation period, $T_2$, for minimising tracking error and maximising the enhanced return over that of the index. In section 5.5, we consider the risk profile of portfolios using the better performing combinations of $T_1$ and $T_2$ by examining three return-risk ratios.

***5.1 Comparison of the pre-selection procedures***

The first part of the analysis compares the results from the pre-selection procedures using 3 years of daily data, $T_1 \approx 754$, for portfolio selection and one year for out-of-sample evaluation, $T_2 \approx 252$. This sequence yields 16 estimations and evaluations and 15 rebalances. For each iteration, the universe consists of the $M$ assets which have always been members of the index over the in-sample period and for which prices are available throughout the duration of the analysis. Over these four-year periods, values of $M$ range from 301 to 427 assets. Primarily, this variation occurs because the constituents of the S&P 500 are adjusted every quarter with around 20 to 30 changes per year mainly due to companies no longer meeting market-capitalisation or financial requirements, mergers and take-overs also have an effect. As all the selection procedures use the same data in each four-year period this does not affect our analyses.

**Figure 3. In-sample tracking errors using BE-OLS (n) for periods starting 2005 to 2020.**

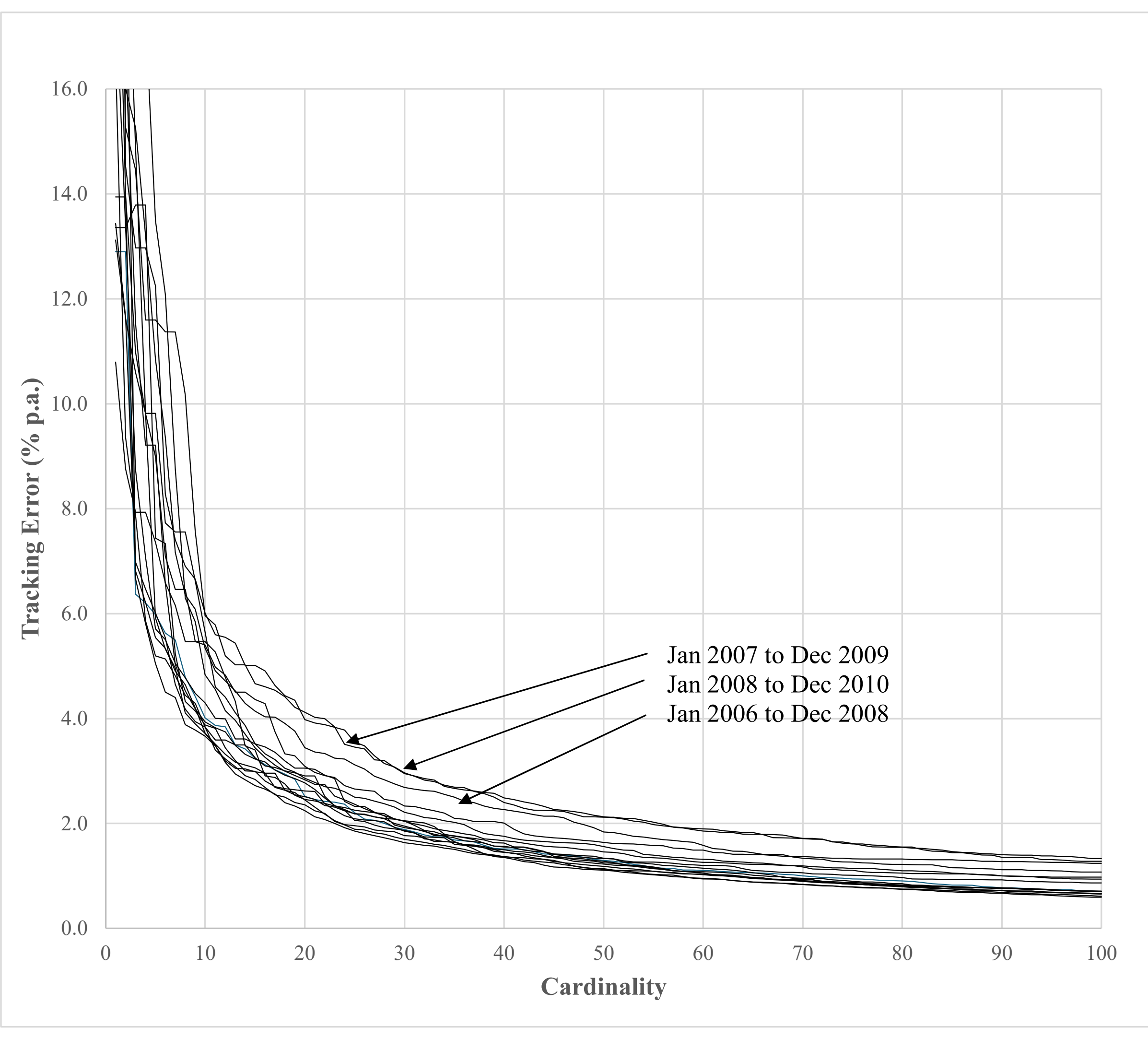

We use each of the pre-selection procedures discussed in Section 3 to populate Δ and calculate the weighting on assets for cardinalities ranging from 1 to 100, minimising the objective function (Equation 6). Example results are shown in Figure 3 for BE-OLS(n), where we show the annualised percentage in-sample tracking error for cardinalities from 1 to 100 for the 16 in-sample periods considered. As expected, there is a dramatic initial drop in tracking error as the cardinality increases.

In Figure 3, the three higher plots show the tracking errors for in-sample periods beginning with 2007, 2008 and 2006, the periods containing the effects of the financial crisis of 2008-9, where the S&P 500 index dropped sharply, and volatility increased (see Figure 2).

To compare the results using each pre-selection procedure, we consider a subset of cardinalities from the 1 to 100 range. In Table 1, we show the average tracking error (% p.a.) in-sample and out-of-sample together with the average transaction volume per annum across the 16 periods. To provide a convenient criterion for comparison, we show the average rank for each measure across cardinalities from 5 to 100, the lower the rank the better. As an example, for in-sample tracking error using FS-OLS(c), we compute its rank among the eight pre-selection procedures. For these 96 cardinalities from 5 to 100, FS-OLS(c) was ranked first once, ranked second 29 times, ranked third 45 times and ranked fourth 21 times, giving the mean rank of 2.9 seen in Table 2.

**Table 1. Summary of the average in- and out-of-sample tracking errors (% p.a.) and transaction volumes p.a. for each pre-selection procedure (using an in-sample of 3 years and an out-of-sample of one year). The average rank (in ascending order) for each category is for all cardinalities from 5 to 100.**

| | | | Average Rank | Cardinality | | | | | | |
|---|---|---|---|---|---|---|---|---|---|---|
| | | | | **5** | **10** | **20** | **30** | **50** | **70** | **100** |
| **With Constant** | **FS-OLS(c)** | **In-sample TE** | 2.9 | 8.41 | 4.75 | 2.97 | 2.20 | 1.53 | 1.19 | 0.94 |
| | | **Out-of-sample TE** | 3.8 | 4.14 | 3.39 | 2.78 | 2.51 | 2.15 | 1.94 | 1.77 |
| | | **Transaction Volume** | 5.2 | 1.69 | 1.63 | 1.53 | 1.44 | 1.36 | 1.25 | 1.11 |
| | **BE-OLS(c)** | **In-sample TE** | 2.9 | 8.07 | 4.86 | 3.03 | 2.25 | 1.52 | 1.17 | 0.89 |
| | | **Out-of-sample TE** | 1.5 | 4.22 | 3.32 | 2.76 | 2.41 | 2.07 | 1.86 | 1.71 |
| | | **Transaction Volume** | 2.9 | 1.55 | 1.46 | 1.43 | 1.40 | 1.24 | 1.13 | 1.01 |
| | **FS-LAD(c)** | **In-sample TE** | 4.8 | 7.62 | 4.78 | 3.23 | 2.32 | 1.69 | 1.33 | 1.04 |
| | | **Out-of-sample TE** | 4.9 | 4.16 | 3.40 | 2.84 | 2.54 | 2.17 | 2.01 | 1.81 |
| | | **Transaction Volume** | 5.0 | 1.55 | 1.52 | 1.56 | 1.45 | 1.34 | 1.26 | 1.15 |
| | **BE-LAD(c)** | **In-sample TE** | 7.6 | 12.60 | 8.07 | 5.37 | 4.41 | 3.36 | 2.83 | 2.36 |
| | | **Out-of-sample TE** | 7.8 | 5.08 | 4.09 | 3.39 | 3.20 | 2.87 | 2.64 | 2.49 |
| | | **Transaction Volume** | 7.4 | 1.96 | 1.94 | 1.90 | 1.86 | 1.74 | 1.67 | 1.64 |
| **Without Constant** | **FS-OLS(n)** | **In-sample TE** | 3.2 | 8.23 | 4.68 | 2.98 | 2.22 | 1.54 | 1.20 | 0.92 |
| | | **Out-of-sample TE** | 3.2 | 4.14 | 3.43 | 2.79 | 2.48 | 2.13 | 1.93 | 1.75 |
| | | **Transaction Volume** | 4.7 | 1.72 | 1.60 | 1.44 | 1.43 | 1.34 | 1.25 | 1.12 |
| | **BE-OLS(n)** | **In-sample TE** | 1.3 | 8.58 | 4.68 | 2.94 | 2.15 | 1.47 | 1.14 | 0.89 |
| | | **Out-of-sample TE** | 1.8 | 4.21 | 3.27 | 2.65 | 2.39 | 2.08 | 1.90 | 1.72 |
| | | **Transaction Volume** | 2.2 | 1.55 | 1.48 | 1.41 | 1.33 | 1.21 | 1.11 | 0.98 |
| | **FS-LAD(n)** | **In-sample TE** | 6.0 | 10.31 | 7.28 | 5.34 | 4.02 | 2.80 | 2.21 | 1.94 |
| | | **Out-of-sample TE** | 6.1 | 4.54 | 3.93 | 3.40 | 2.99 | 2.54 | 2.28 | 2.21 |
| | | **Transaction Volume** | 1.0 | 0.61 | 0.60 | 0.48 | 0.44 | 0.51 | 0.44 | 0.49 |
| | **BE-LAD(n)** | **In-sample TE** | 7.4 | 12.49 | 8.83 | 5.68 | 4.46 | 3.38 | 2.76 | 2.27 |
| | | **Out-of-sample TE** | 7.1 | 4.90 | 4.08 | 3.42 | 3.15 | 2.77 | 2.55 | 2.39 |
| | | **Transaction Volume** | 7.6 | 1.83 | 1.91 | 1.86 | 1.88 | 1.74 | 1.71 | 1.65 |

BE-OLS(n) ranks highly across the three measures and is closely followed by the other OLS procedures, BE-OLS(c), FE-OLS(c) and FS-OLS(n). Of the LAD based procedures, forward selection is noticeably better than backward elimination. An explanation for the relatively poor performance of BE-LAD(c or n) may be that in the early stages of the selection of assets, the large number of assets in the solutions leads to degeneracy in the linear programs underlying LAD regression. FS-LAD(n) shows the lowest transaction volume, indicating comparatively fewer changes in the selected assets over the 16 periods, but the resultant in-sample tracking error is poor.

An unexpected property of the tracking errors for the lower valued cardinalities is that the out-of-sample tracking error is less than the in-sample tracking error. A priori, one would expect fitted weights in-sample to capture the 'signal' of the index, out-of-sample one would expect the 'noise' around the 'signal' to cause the tracking error to deteriorate. This is what happens for the higher cardinalities (for cardinalities of 30 or more using the OLS selection procedures). The lower cardinalities are less good at capturing the 'signal' and their tracking errors are dominated by 'noise'. Comparing the volatility of the returns on the S&P 500 over a 3-year in-sample period with the volatility over the following 1-year out-of-sample period, shows that the in-sample volatility is greater in 11 out of our 16 comparisons, and more than 30% greater for 6 of the comparisons. In other words, for the lower cardinalities, the smaller out-of-sample tracking errors are reflecting that, on average, there is less 'noise' out-of-sample.

In Table 1, we see that the tracking errors, both in-sample and out-of-sample, from the four OLS selection procedures are similar. To judge the similarity of the choices of assets made by these procedures, we look at the intersections of the sets of assets chosen in Table 2. We give the average percentage of assets in common to the tracking portfolios from the different selection procedures across a range of cardinalities. For example, for BE-OLS(c) vs BE-OLS(n) with a cardinality of 20, on average 60.6% of the assets are common to the portfolios chosen by these procedures.

**Table 2. Summary of the percentage of assets common to the choices resulting from some pairs of different pre-selection procedures**

| | **Cardinality** | | | | | | |
|---|---|---|---|---|---|---|---|
| | **5** | **10** | **20** | **30** | **50** | **70** | **100** |
| BE-OLS(c) vs FS-OLS(c) | 21.3 | 23.8 | 25.3 | 29.4 | 35.6 | 47.1 | 60.3 |
| BE-OLS(c) vs BE-OLS(n) | 55.0 | 55.0 | 60.6 | 63.1 | 71.9 | 78.5 | 83.3 |
| FS-LAD(c) vs FS-OLS(c) | 33.8 | 27.5 | 25.3 | 29.2 | 33.5 | 41.5 | 49.4 |
| FS-LAD(c) vs BE-LAD(c) | 1.3 | 1.9 | 6.3 | 8.3 | 12.0 | 17.8 | 25.0 |

The effect of the omission of the constant in the OLS regression is shown with BE-OLS(c) vs BE-OLS(n) where the proportion of assets in common increases from a half to over 80% as the cardinality increases. The difference in proportion of assets in common between forward selection and backward elimination, BE-OLS(c) vs FS-OLS(c), shows a greater difference, less than 50% of assets in common for cardinalities of 70 or less. The difference in proportion of assets in common between OLS and LAD, FS-LAD(c) vs FS-OLS(c), is also large with less than 50% of assets in common for all cardinalities. The last row in Table 2, FS-LAD(c) vs BE-LAD(c), demonstrates the contrast between BE-LAD and the other procedures.

Finding a unique optimal solution that minimises in-sample tracking error does not guarantee an optimal out-of-sample tracking error. The concept of a single optimal tracking portfolio is further undermined in Table 2, where the constituents in common to similarly performing tracking portfolios can be very different. This difference in constituents is particularly evident in the comparison between BE-OLS(c) and FS-OLS(c) where for a wide range of cardinalities, two sets of nearly equally well performing tracking portfolios share less than half the same assets.

### *5.2 Explanation of variation in tracking error over time*

From Figure 3, we see that the in-sample tracking error at different cardinalities varies according to the period used for estimation, as we would expect. In Table 3, we compare the out-of-sample tracking error for BE-OLS(n) for a selection of cardinalities and compare the tracking errors with the volatility of the index during the out-of-sample period.

**Table 3. Out-of-sample tracking error (% p.a.) for BE-OLS(n) for a selection of cardinalities compared with index volatility (% p.a.) during years 2008 to 2023.**

| | | Cardinality | | | | | | |
|---|---|---|---|---|---|---|---|---|
| **Date** | **Volatility** | **5** | **10** | **20** | **30** | **50** | **70** | **100** |
| 2005 | 10.25 | | | | | | | |
| 2006 | 9.88 | | | | | | | |
| 2007 | 15.99 | | | | | | | |
| 2008 | 41.05 | 11.91 | 6.24 | 4.90 | 4.40 | 3.87 | 3.20 | 2.57 |
| 2009 | 27.24 | 7.05 | 4.75 | 3.94 | 3.38 | 3.01 | 2.15 | 1.84 |
| 2010 | 18.06 | 4.38 | 3.10 | 2.52 | 2.24 | 2.01 | 1.58 | 1.37 |
| 2011 | 23.33 | 4.85 | 3.60 | 3.16 | 2.71 | 2.47 | 1.86 | 1.73 |
| 2012 | 12.86 | 4.71 | 2.64 | 2.43 | 1.96 | 1.79 | 1.36 | 1.24 |
| 2013 | 11.07 | 4.81 | 3.13 | 2.67 | 1.95 | 1.75 | 1.29 | 1.12 |
| 2014 | 11.33 | 3.85 | 3.16 | 2.22 | 1.93 | 1.68 | 1.20 | 1.10 |
| 2015 | 15.48 | 5.17 | 3.07 | 2.46 | 1.97 | 1.84 | 1.37 | 1.22 |
| 2016 | 13.02 | 4.58 | 3.15 | 2.74 | 2.28 | 1.80 | 1.50 | 1.27 |
| 2017 | 6.67 | 3.70 | 3.04 | 2.29 | 1.91 | 1.62 | 1.21 | 1.06 |
| 2018 | 17.04 | 4.35 | 3.07 | 2.61 | 2.20 | 1.91 | 1.35 | 1.06 |
| 2019 | 12.48 | 3.24 | 2.44 | 2.08 | 1.71 | 1.58 | 1.31 | 1.13 |
| 2020 | 34.74 | 6.51 | 4.50 | 3.52 | 3.03 | 2.35 | 1.59 | 1.45 |
| 2021 | 13.10 | 4.60 | 3.18 | 2.57 | 2.37 | 2.23 | 1.74 | 1.71 |
| 2022 | 24.14 | 6.02 | 3.72 | 2.98 | 2.48 | 2.28 | 1.67 | 1.62 |
| 2023 | 13.06 | 5.65 | 3.58 | 2.77 | 2.56 | 2.34 | 2.18 | 2.13 |
| Range | 34.38 | 8.67 | 3.80 | 2.82 | 2.69 | 2.30 | 2.00 | 1.52 |
| Correlation | | 0.85 | 0.89 | 0.91 | 0.90 | 0.86 | 0.74 | 0.67 |

Looking down the columns, we see wide variations over the different out-of-sample years, measuring this variation by the range of values, we see that the range decreases as cardinality increases. Further, we see a strong correlation (significant at 0.5% or less) between the out-of-sample tracking error in a year and the index volatility. This correlation tends to decrease as the cardinality increases. These results indicate that increasing the cardinality reduces the portfolio's dependence on index volatility, in addition to improving out-of-sample stability.

### *5.3 Modelling the relation between tracking error and cardinality*

For the rest of our study, we focus on OLS selection procedures as Table 1 shows they are more effective than the LAD procedures The relationship between tracking error and cardinality ($n$) is investigated using the following model which controls for the selection procedure used. The dummy variables $I_{procedure}$ are set to 1 for tracking errors produced by that procedure, 0 otherwise, the base procedure is BE-OLS(n). The regression equation used is:

$$ln(TE_n) = \alpha_0 + \alpha_1 ln(n) + \alpha_2 I_{FS-OLS\,(c)} + \alpha_3 I_{FS-OLS\,(n)} + \alpha_4 I_{BE-OLS\,(c)} \quad (7)$$

The model is estimated separately for in- and out-of-sample tracking errors. The results are shown in Table 4.

**Table 4. The relationship between tracking error and cardinality (*n*), estimates of the coefficients in Equation (7).**

| | **In-sample** | *p-value* | **Out-of-sample** | *p-value* |
|---|---|---|---|---|
| Number of obs. | 6144 | | 6144 | |
| Adjusted R$^2$ | 0.85 | | 0.73 | |
| $\alpha_0$ | 3.21 | *0.00* | 2.99 | *0.00* |
| $\alpha_1 ln(n)$ | -0.73 | *0.00* | -0.58 | *0.00* |
| $\alpha_2$FS-OLS(c) | 0.03 | *0.00* | 0.06 | *0.00* |
| $\alpha_3$FS-OLS(n) | 0.03 | *0.00* | 0.05 | *0.00* |
| $\alpha_4$BE-OLS(c) | 0.02 | *0.01* | 0.00 | *0.94* |

The adjusted R$^2$ s of the estimations show that the log – log regression is a plausible fit, it is easier to interpret the results by taking anti-logs:

$$TE = {}^{\theta}/_{n^{\omega}} + \epsilon \quad (8)$$

where $\theta$ is the notional base-line tracking error for a cardinality of one asset, the tracking error for increasing cardinalities decreases with $n$ to the power of $\omega$, the greater the value of $\omega$, the faster the decline in tracking error. For the in-sample tracking error of the BE-OLS(n) procedure, $\omega = 0.73$ and $\theta = e^{3.21} = 24.8$, the other procedures produce significantly higher tracking errors (by a factor of exp($\alpha_i$) where $i$ = 2, 3 or 4). Out-of-sample, the two forward selection procedures produce significantly higher tracking errors, but there is no significant difference between BE-OLS with or without a constant.

For the out-of-sample tracking error of the BE-OLS(n) procedure, $\omega = 0.58$ and $\theta = e^{2.99}$. Rounding down 0.58 to 0.5, we have the intuitively convenient interpretation of (8) that out-of-sample tracking error is roughly proportional to $1/\sqrt{cardinality}$.

### *5.4 Enhanced index tracking and sensitivity analysis*

For our analysis so far, we have used three years of daily data for asset selection and optimisation of asset weights and one year for out-of-sample evaluation ($T_1 \approx 754,\ T_2 \approx 252$) for tracking error of the index. In this section, we investigate tracking an enhanced index return by setting λ to the equivalent of *r*% p.a., where (*r* = 0,1,2,3,4,5). We consider three estimation periods of 2, 3 and 4 years ($T_1 \approx 504, 754, 1008$ ) and three evaluation periods of a year, 6 months and 3 months ($T_2 \approx 252, 126, 63$). For $T_1 \approx 754, 1008$ we use BE-OLS(n), however since BE-OLS(n) needs $T_1 > M$, for the shorter estimation periods FS-OLS(n) is used for asset selection.

We consider two sensitivity analyses: firstly, we consider out-of-sample tracking error; secondly, we consider out-of-sample enhanced index return.

**Table 5. Sensitivity analysis for a tracking portfolio: a summary of sensitivity to the choice of $T_1$ and $T_2$ on out-of-sample tracking error (% p.a.) and transaction volume using FS-OLS(n) for $T_1$ = 2 years and using BE-OLS(n) for $T_1$ = 3 and 4 years. The average rank is calculated over all cardinalities from 5 to 100.**

| | | | | **Cardinality** | | | | | | |
|---|---|---|---|---|---|---|---|---|---|---|
| **T₁** | **T₂** | | **Average rank** | **5** | **10** | **20** | **30** | **50** | **70** | **100** |
| | 3 months | Tracking Error | 6.9 | 4.05 | 3.35 | 2.77 | 2.46 | 2.12 | 1.92 | 1.72 |
| | | *Transaction Volume* | *8.9* | *5.47* | *5.82* | *5.82* | *5.73* | *5.37* | *4.84* | *4.28* |
| 2 years | 6 months | Tracking Error | 8.0 | 4.04 | 3.34 | 2.79 | 2.49 | 2.13 | 1.93 | 1.76 |
| | | *Transaction Volume* | *6.0* | *3.31* | *3.24* | *3.15* | *3.07* | *2.81* | *2.59* | *2.28* |
| | 1 year | Tracking Error | 8.9 | 4.03 | 3.37 | 2.81 | 2.53 | 2.16 | 1.95 | 1.76 |
| | | *Transaction Volume* | *3.0* | *1.83* | *1.75* | *1.63* | *1.58* | *1.45* | *1.36* | *1.23* |
| | 3 months | Tracking Error | 3.1 | 4.08 | 3.28 | 2.66 | 2.36 | 2.02 | 1.83 | 1.64 |
| | | *Transaction Volume* | *8.0* | *5.80* | *5.64* | *4.94* | *4.58* | *3.86* | *3.39* | *2.89* |
| 3 years | 6 months | Tracking Error | 4.1 | 4.25 | 3.26 | 2.66 | 2.37 | 2.04 | 1.84 | 1.68 |
| | | *Transaction Volume* | *5.0* | *3.37* | *3.07* | *2.64* | *2.50* | *2.17* | *1.98* | *1.70* |
| | 1 year | Tracking Error | 5.9 | 4.21 | 3.27 | 2.65 | 2.39 | 2.08 | 1.90 | 1.72 |
| | | *Transaction Volume* | *1.9* | *1.55* | *1.48* | *1.41* | *1.33* | *1.21* | *1.11* | *0.98* |
| | 3 months | Tracking Error | 1.1 | 4.09 | 3.21 | 2.60 | 2.29 | 1.96 | 1.78 | 1.61 |
| | | *Transaction Volume* | *7.1* | *6.15* | *5.51* | *4.89* | *4.25* | *3.52* | *3.00* | *2.39* |
| 4 years | 6 months | Tracking Error | 2.6 | 4.12 | 3.34 | 2.64 | 2.33 | 2.01 | 1.82 | 1.64 |
| | | *Transaction Volume* | *4.0* | *3.10* | *2.99* | *2.57* | *2.28* | *1.97* | *1.75* | *1.42* |
| | 1 year | Tracking Error | 4.5 | 3.92 | 3.31 | 2.61 | 2.39 | 2.06 | 1.88 | 1.68 |
| | | *Transaction Volume* | *1.1* | *1.56* | *1.57* | *1.38* | *1.27* | *1.12* | *0.98* | *0.89* |

The results of different choices of $T_1$ and $T_2$ on out-of-sample tracking error and the associated transaction volume are summarised in Table 5. In this table the rank values are calculated as described above for Table 1, where the lower the average rank the better. Looking first at the choice of the estimation period, $T_1$, we see that tracking error tends to decrease as $T_1$ increases, similarly the transaction volume also decreases as $T_1$ increases. Considering the choice of the evaluation period, $T_2$, within each value of $T_1$, the tracking error decreases as the evaluation period shortens. Concomitantly, the transaction volume increases as the evaluation period halves, because the number of rebalances doubles. The combination that offers the lowest tracking error at most cardinalities is an in-sample period of 4 years coupled with an out-of-sample period of 3 months ($T_1 \approx 1008$, $T_2 \approx 63$), however the combination that offers the lowest transaction volume is an in-sample period of 4 years coupled with an out-of-sample period of one year ($T_1 \approx 1008$, $T_2 \approx 252$). Depending on the relative importance of these measures, a compromise might be an in-sample period of 4 years coupled with an out-of-sample period of 6 months ($T_1 \approx 1008$, $T_2 \approx 126$) where both the tracking error and the transaction volume increase slightly. However, the most important choice is that of cardinality, Table 5 clearly shows that the larger the number of assets in the portfolio, the lower the tracking error and transaction volume.

**Table 6. Sensitivity analysis for an enhanced tracker: a summary of sensitivity to the choice of $T_1$ and $T_2$ on out-of-sample excess return (% p.a.) over the index and transaction volume using FS-OLS(n) for $T_1$ = 2 years and using BE-OLS(n) for $T_1$ = 3 and 4 years. The average rank is calculated over all cardinalities from 5 to 100.**

| | | | | **Cardinality** | | | | | | |
|---|---|---|---|---|---|---|---|---|---|---|
| **$T_1$** | **$T_2$** | | **Average rank** | **5** | **10** | **20** | **30** | **50** | **70** | **100** |
| 2 years | 3 months | Tracking Error | 3.8 | 0.58 | 2.01 | 2.25 | 1.64 | 0.88 | 0.80 | 0.61 |
| | | *Transaction Volume* | *8.9* | *5.17* | *5.46* | *5.65* | *5.57* | *5.21* | *4.78* | *4.31* |
| | 6 months | Tracking Error | 4.9 | -1.04 | 1.07 | 1.45 | 1.42 | 0.75 | 0.35 | 0.65 |
| | | *Transaction Volume* | *6.0* | *3.19* | *3.15* | *3.07* | *2.98* | *2.76* | *2.54* | *2.31* |
| | 1 year | Tracking Error | 8.7 | -0.77 | -0.41 | 0.56 | 0.09 | 0.45 | -0.03 | 0.26 |
| | | *Transaction Volume* | *3.0* | *1.83* | *1.73* | *1.69* | *1.62* | *1.46* | *1.38* | *1.24* |
| 3 years | 3 months | Tracking Error | 4.1 | 0.53 | 1.17 | 1.42 | 0.96 | 0.97 | 0.91 | 1.07 |
| | | *Transaction Volume* | *8.0* | *5.76* | *5.31* | *4.75* | *4.28* | *3.83* | *3.38* | *2.89* |
| | 6 months | Tracking Error | 1.6 | -1.44 | 2.77 | 1.58 | 1.66 | 1.57 | 1.45 | 1.26 |
| | | *Transaction Volume* | *5.0* | *3.15* | *2.87* | *2.63* | *2.48* | *2.15* | *1.96* | *1.76* |
| | 1 year | Tracking Error | 5.5 | 0.49 | 0.60 | -0.13 | 0.32 | 0.99 | 0.74 | 0.78 |
| | | *Transaction Volume* | *1.9* | *1.62* | *1.48* | *1.41* | *1.32* | *1.21* | *1.09* | *1.00* |
| 4 years | 3 months | Tracking Error | 3.9 | 1.26 | 0.22 | 0.81 | 1.16 | 0.79 | 1.03 | 1.10 |
| | | *Transaction Volume* | 7.0 | *5.34* | *4.91* | *4.34* | *4.12* | *3.41* | *2.95* | *2.53* |
| | 6 months | Tracking Error | 5.4 | 0.72 | 0.13 | 0.99 | 1.24 | 0.58 | 0.30 | 0.73 |
| | | *Transaction Volume* | *4.0* | *2.65* | *2.42* | *2.32* | *2.16* | *2.00* | *1.75* | *1.53* |
| | 1 year | Tracking Error | 7.2 | 1.15 | 1.06 | 0.32 | 0.62 | 0.18 | 0.36 | 0.30 |
| | | *Transaction Volume* | *1.1* | *1.50* | *1.40* | *1.33* | *1.25* | *1.17* | *1.08* | *0.93* |

We now focus on the out-of-sample enhanced return, where the objective is to track the return of the index plus an enhancement, in this case the desired enhancement λ was set to the equivalent of 5% p.a. The results of different choices of $T_1$ and $T_2$ are summarised in Table 6, along with the associated transaction volume. Although the objective of an enhanced return of 5% was achieved in-sample, this enhancement deteriorated out-of-sample in all cases. In Table 5 the optimum is associated with the longest estimation period, the shortest evaluation period and the largest cardinality of the values considered. In Table 6 the picture is more complicated, the combination offering the highest enhanced return at most cardinalities is an in-sample period of 3 years coupled with an out-of-sample period of 6 months ($T_1 \approx 754$, $T_2 \approx 126$). In Figure 4, we show the enhanced return over the index for the two higher ranked combinations of $T_1$ and $T_2$. For an in-sample period of 3 years coupled with an out-of-sample period of 6 months, the higher enhanced returns result from cardinalities of 9 and 10 and then tail off as the cardinality is increased. The pattern is similar for an in-sample period of 2 years coupled with an out-of-sample period of 3 months ($T_1 \approx 504$, $T_2 = 63$) where the higher enhanced returns occur at cardinalities of 19 and 20 and then tail off. The problem of choosing a portfolio to track an index with an enhanced return is a balancing act between identifying outperforming assets in the estimation period and choosing an evaluation period where the outperformance persists. Table 6 and Figure 4 demonstrate that the balance is found with portfolios containing relatively few assets.

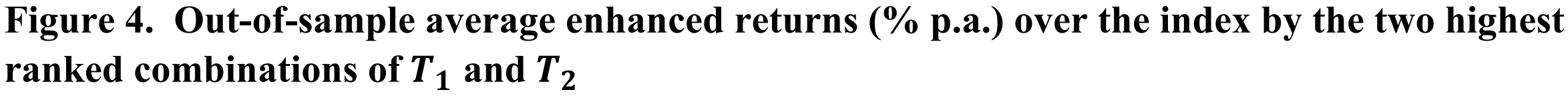

**Figure 4. Out-of-sample average enhanced returns (% p.a.) over the index by the two highest ranked combinations of $T_1$ and $T_2$**

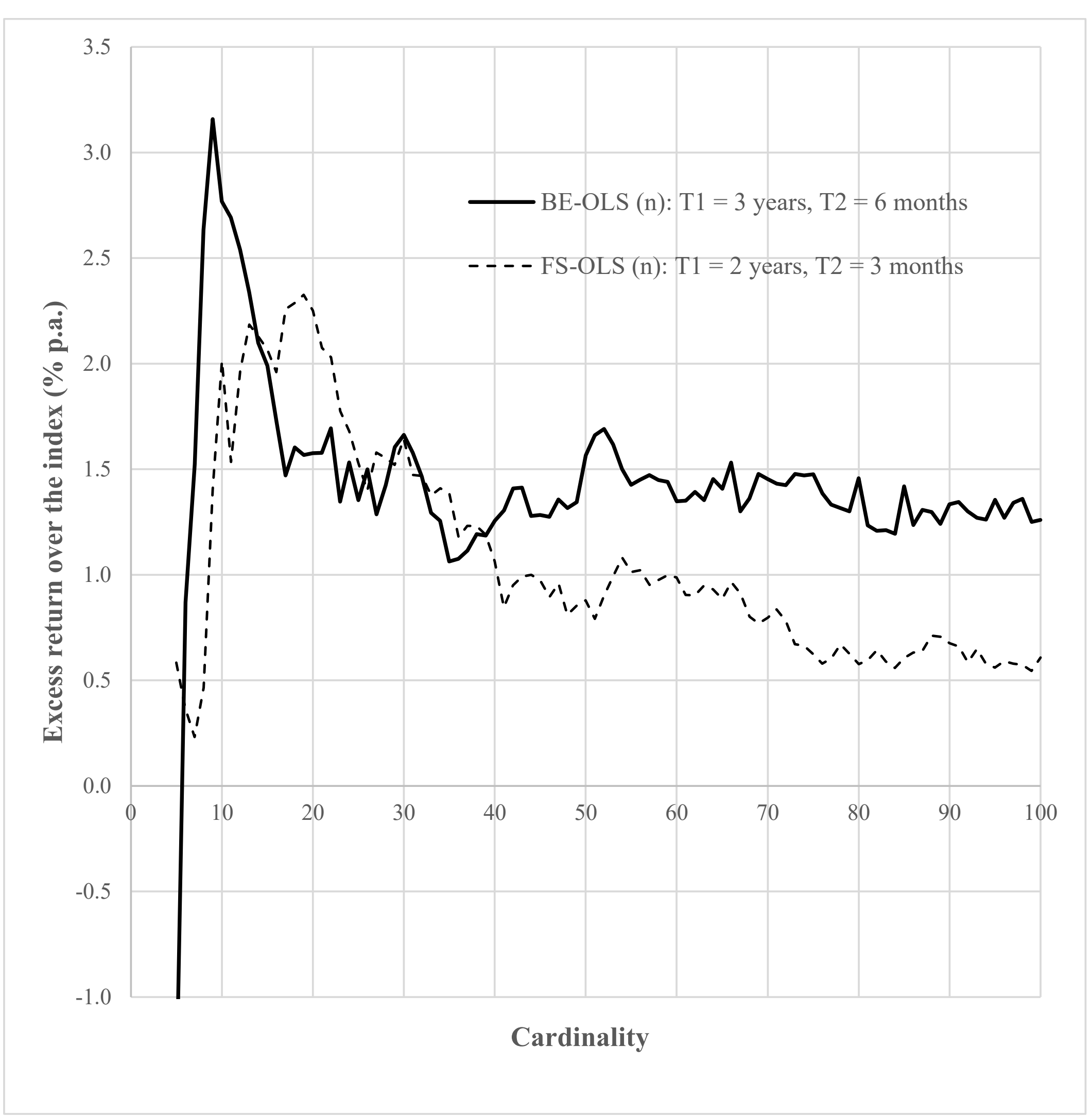

### *5.5 Return-risk ratios for the tracking portfolios*

Return-risk ratios are a conventional way of comparing portfolio performance. We consider three return-risk ratios, if X is the excess return i.e. (the return on a portfolio - the risk-free rate) then:

the Sharpe ratio is $$E(X)\big/SD(X) \quad (9)$$

the Gains-Loss ratio is $$E(X|X>0)\big/|E(X|X\leq 0)| \quad (10)$$

the Sortino ratio is $$E(X|X>0)\big/SD(X|X\leq 0) \quad (11).$$

For all these ratios, the higher the values the better, the relative merits of these return-risk ratios are discussed in Cheridito and Kromer (2013) and details of the calculation of these ratios are given in the Appendix.

The values of these return-risk ratios, calculated for the best performing combination of $T_1$ and $T_2$ for minimising tracking error and for maximising out-of-sample enhanced returns, are shown for a range of cardinalities in Table 7. For the best tracker, we see that all three ratios trend upwards as cardinality increases, at a cardinality of 100 the ratio values correspond closely with the values of the index itself. For the best enhancement portfolios, the ratios are all at their highest value for a cardinality of 10 and decrease gradually as the cardinality increases. Note that these values are all smaller than their corresponding values for the best tracking portfolios, this is because that although on average the returns are higher, the risks are also higher.

**Table 7. Return-risk ratios for the best $T_1$ and $T_2$ combinations for minimum tracking error and maximum enhancement over a range of cardinalities.**

| **Best tracker: BE-OLS(n); $T_1$ = 4 years; $T_2$ = 3 months** | | | | | |
|---|---|---|---|---|---|
| Cardinality | Tracking Error | Transaction Volume | Sharpe | Gain-loss | Sortino |
| 10 | 3.21 | 5.51 | 0.418 | 1.081 | 0.582 |
| 20 | 2.60 | 4.89 | 0.429 | 1.085 | 0.594 |
| 30 | 2.29 | 4.25 | 0.466 | 1.093 | 0.645 |
| 40 | 2.09 | 3.85 | 0.442 | 1.088 | 0.610 |
| 50 | 1.96 | 3.52 | 0.455 | 1.091 | 0.627 |
| 60 | 1.86 | 3.20 | 0.458 | 1.091 | 0.633 |
| 70 | 1.78 | 3.00 | 0.452 | 1.090 | 0.625 |
| 80 | 1.72 | 2.78 | 0.467 | 1.093 | 0.645 |
| 90 | 1.66 | 2.55 | 0.472 | 1.094 | 0.652 |
| 100 | 1.61 | 2.39 | 0.474 | 1.095 | 0.654 |
| Index | | | 0.474 | 1.095 | 0.653 |
| **Best Enhancement: BE-OLS(n); $T_1$ = 3 years; $T_2$= 6 months** | | | | | |
| Cardinality | Enhanced return | Transaction Volume | Sharpe | Gain-loss | Sortino |
| 10 | 2.77 | 2.87 | 0.363 | 1.073 | 0.504 |
| 20 | 1.58 | 2.63 | 0.312 | 1.063 | 0.431 |
| 30 | 1.66 | 2.48 | 0.318 | 1.064 | 0.439 |
| 40 | 1.26 | 2.28 | 0.299 | 1.060 | 0.412 |
| 50 | 1.57 | 2.15 | 0.314 | 1.064 | 0.432 |
| 60 | 1.35 | 2.04 | 0.304 | 1.061 | 0.417 |
| 70 | 1.45 | 1.96 | 0.310 | 1.063 | 0.425 |
| 80 | 1.46 | 1.88 | 0.309 | 1.063 | 0.424 |
| 90 | 1.33 | 1.78 | 0.304 | 1.062 | 0.417 |
| 100 | 1.26 | 1.76 | 0.300 | 1.061 | 0.411 |

## 6. Summary and Conclusions

In our approach, we divide the problem of selecting assets for a CCITP into two steps: pre-selection of assets for a portfolio of given cardinality; followed by calculation of weights on the assets chosen. Our approach allows cardinalities of any size to be considered giving the portfolio manager important flexibility. We show that the sharp reduction in out-of-sample tracking error as cardinality increases is roughly proportional to $1/\sqrt{cardinality}$ .

We investigated eight pre-selection procedures for the assets to be chosen for index tracking portfolios of a given cardinality. These two selection approaches, forward selection and backward elimination, were implemented using OLS and LAD regression and each regression was considered with and without a constant term. We found that OLS procedures dominated LAD procedures and that BE-OLS without a constant was marginally superior. We demonstrate that the out-of-sample tracking error is highly correlated with the contemporary volatility of the underlying market.

The difficulty of establishing a unique optimal solution to the selection of a cardinality constrained index tracking portfolio is highlighted by our analysis showing that the proportion of assets common to similarly performing, locally optimal, tracking portfolios (e.g. BE-OLS(c) and FS-OLS(c)) is less than half in nearly all comparisons. In other words, the multivariate surface defined by the asset weights is very flat near the optimum in-sample tracking error.

In our sensitivity analysis relating to the length of estimation period and the length of the evaluation period, we introduced portfolios tracking an enhanced index, i.e. offering the index return plus an enhancement. Regarding building index tracking portfolios, the sensitivity analysis confirmed the picture emerging from the initial analyses, tracking error and transaction volume both decrease as cardinality increases. In other words, the less effect the cardinality constraint exerts, the better the tracking performance. However, the benefits of cardinality constraints are restored when building an enhanced tracking portfolio, where we showed that cardinalities below 20 offered better enhanced returns.

Intuitively, the index tracker is seeking an 'average' performance as represented by the index which is achieved by including as many assets as available; we show that the tracking portfolio with estimation period of 4 years and an evaluation period of 3 months exhibits the same return-risk ratios as the index itself. In contrast the enhanced tracking portfolio needs to identify the few assets that outperform the index by a given margin. The pre-selection during the estimation period identifies these outperforming assets and the assumption is that the momentum of these assets persists, at least in part, during the evaluation period. The sensitivity analysis suggests that an estimation period of 3 years coupled with an evaluation period of 6 months is the most effective. Possible developments of this approach include: further investigation into the pre-selection of assets perhaps using machine learning techniques, such as random forests for selection from large universes such as the Russell indices; here we have looked at an additive enhancement, exploration of different forms, such as multiplicative enhancement, could be considered.

**Appendix. Calculation of Reward-Risk ratios**

The formulae in equations (6), (7) and (8) are in terms of expected values, here we describe their estimation from data, namely $T$ observations of daily index excess returns or daily out-of-sample excess returns from January 2008 to December 2023. The risk-free rate used is $R_t$, the CBOE 10-year US Treasury note (in % p.a.) (symbol TNX), which means that the daily risk-free rate is $rf_t = (1 + 0.01R_t)^{1/252} - 1$. Excess return means:

for the index $y_t = \left(\ln\left({}^{I_t}/_{I_{t-1}}\right) - rf_t\right)$

or for a tracking portfolio $y_t = \left(\ln\left({}^{TP_{nt}}/_{TP_{n,t-1}}\right) - rf_t\right)$.

Let $S_1 = \sum_{t=1}^{T} y_t$, $S_2 = \sum_{t=1}^{T} y_t^2$, $S_1^+ = \sum_{t=1}^{T}(y_t | y_t > 0)$, $S_1^- = -\sum_{t=1}^{T}(y_t | y_t \leq 0)$ and $S_2^- = \sum_{t=1}^{T}(y_t^2 | y_t \leq 0)$.

For the Sharpe ratio, the expected annual return is $mean = 100 \times 252 \times \left({}^{S_1}/_T\right)$, the annualised standard deviation is $SD = 100 \times \sqrt{252 \times {}^{\left(S_2 - {}^{S_1^2}/_T\right)}/_T}$ and the ratio is $\left({}^{mean}/_{SD}\right)$.

The Gains/Loss ratio is ${}^{S_1^+}/_{S_1^-}$.

The Sortino ratio is annualised here (for discussion of calculation details, see Kidd (2012)), the downside risk is $SD^- = 100 \times \sqrt{252 \times {}^{S_2^-}/_T}$; the Sortino ratio is ${}^{mean}/_{SD^-}$ .